\documentstyle[amssymb,epsfig,prb,aps,multicol]{revtex}
\begin{document}
\title{Continuous phase transition in polydisperse hard-sphere mixture}
\author{Ronald Blaak and Jos\'e A.~Cuesta}
\address{Grupo Interdisciplinar de Sistemas Complicados (GISC),
Dpto.~de Matem\'aticas, Univ.~Carlos III de Madrid, Avda.~de la
Universidad 30, 28911 Legan\'es, Spain} 
\date{\today}
\maketitle

\begin{abstract}
In a previous paper (J. Zhang {\it et al.}, J. Chem. Phys. {\bf 110},
5318 (1999)) we introduced a model for polydisperse hard sphere
mixtures that is able to adjust its particle-size distribution. Here
we give the explanation of the questions that arose in the previous
description and present a consistent theory of the phase transition in
this system, based on the Percus-Yevick equation of state. The
transition is continuous, and like Bose-Einstein condensation a
macroscopic aggregate is formed due to the microscopic interactions. A
BMCSL-like treatment leads to the same conclusion with slightly more
accurate predictions. 
\end{abstract}

\begin{multicols}{2}

\section{Introduction}
\label{sec:intro}
Monodisperse colloidal suspensions are rare in nature. Whether these
colloidal particles are artificially prepared or not, in the best case
the particle-size distribution is a narrow distribution around an
average size. It is obvious that such a polydisperse nature will
influence the physics and properties of these systems, and one could
imagine that by the use of truly asymmetric mixtures it is possible to
create systems with a behavior that cannot be described by the simple
monodisperse-like approximation. 

Conceptually we can distinguish between two types of polydispersity.
One of them, which we could refer to as `intrinsic polydispersity',
arises from the fact that the particles present in the system are
different by construction (in size, charge, or any other feature)
and their characteristics are not changed
by the interaction with other particles. This kind of systems are
like multicomponent mixtures in which, at least in principle, the
composition can be externally imposed. The new phenomenology that
we can expect from this systems has its origin in fractionations into
phases with different compositions \cite{Sear:98EPL,Bartlett:98JCP,%
Cuesta:99EPL,Warren:99EPL} (constrained by particle conservation)
and their coupling with other transitions already present in the
monodisperse system \cite{Bartlett:99PRL,Bates:1998JCP}.

The second kind of polydispersity can
be found in self-assembling systems \cite{Gelbart:1994book}
(surfactants forming micelles,
monomers forming chains, vesicles, etc). The aggregates present
in these systems can be identified as the particles, each with 
different size, shape, conformation, etc; the difference with the
intrinsically polydisperse systems being that the composition is
determined by the chemical equilibrium between the constituents of
the aggregates. As a consequence the equilibrium is not constrained by
conservation of the number of particles and therefore no fractionation
is to be expected. In principle 
the system can compensate losses of entropy by adjusting its
composition. There are, however, other constraints in the system
(the number of small constituents, for instance) and these may
induce new kinds of transitions characterized by the appearance of
one or a few macroscopic aggregates. As such can be cataloged phenomena
 as the appearance of lamellar and columnar phases in 
surfactant solutions \cite{Boden:1986CPL,Boden:1987JPC}, 
emulsification failure in micro-emulsions \cite{Leaver:1994JPII,%
Vollmer:1997JCP} or long chain formation in polymer
solutions \cite{Petschek:1986PRA}.

A study of polydisperse mixtures of either type is far 
from trivial. In the case of intrinsic polydispersity there is the 
experimental problem of how to fabricate colloidal particles according 
to a given particle-size distribution. Although in simulations this
seems to be somewhat better under control, one easily could run into
the problem of finite size effects due to an insufficient or
inadequate sampling of particle sizes. This is not the case in
self-assembling systems, although experimentally, their polydispersity 
cannot be easily characterized. 

Theoretical descriptions of polydisperse systems are mainly
based on a small set of moments of the particle-size distribution
\cite{Cuesta:99EPL,Warren:99EPL,Sollich:1998PRL,Warren:1998PRL,%
Sollich:2001ACP,Clarke:2000JCP} and
it is assumed that mixtures with the same set of moments show a
corresponding behavior. This seems to be a rather successful
approach, although it is not obvious that could still be applied to 
very asymmetric mixtures. 

In two previous articles we analyzed the behavior of idealized
versions of polydisperse mixtures. The system was assumed to be
composed of $N$ spherical aggregates which only interact via hard-core 
repulsion. The chemical equilibrium of the underlying constituents 
is accounted for by allowing particles to exchange size in such a way 
that the total volume \cite{Zhang:1999JCP} (compact aggregates) or 
surface \cite{Blaak:2000JCP} (surfactant micellar membranes) of 
the particles remains fixed at all times and the number of particles 
is constant.  

Under the influence of the applied pressure the particle-size
distribution of these systems changes. In the case of the constrained
surface, the particle-size distribution changed from an exponentially
decaying function in the low density limit to a single peaked
distribution in the denser liquid phase. For sufficiently high
pressures this system can form a number of different mechanically
stable crystals, a monodisperse face-centered-cubic crystal, or a
bidisperse $AB$ or $AB_2$ crystal.  

For the system with the constraint on the total volume of the
particles, it was found by theory and simulations that at a rather low
volume fraction of approximately  $0.26$ the system could no longer be
described by Percus-Yevick type of equation of state, due to the
formation of macroscopically large particles. In the present work we
will show that this phenomenon is a true phase transition and provide
a self-consistent theory for this sort of system. The nature of this
transition has recently been studied in ideal systems \cite{Sear:2001}
and found
to be connected to Bose-Einstein condensation, and from this work
we can conclude that interaction only changes the details, not the
essential features. 
The connection of the present model with  Bose-Einstein
condensation was earlier suggested by D.~Frenkel \cite{Frenkel}. 

The remainder of this paper is organized as follows. In
Sec.~\ref{sec:MC} we will show the results of computer simulations we
have performed, and explain some of the details involved. These
results are compared with a theory based on the Percus-Yevick equation
of state in Sec.~\ref{sec:PY}. As we show in Sec.~\ref{sec:BMCSL} the
treatment based on the heuristic BMCSL equation of state, leads to a
slightly better theoretical description. In Sec.~\ref{sec:discussion}
we finish with a brief discussion on some of the issues raised in this
paper. 

\section{Simulations}
\label{sec:MC}
The system under consideration here is one formed by $N$ spherical
particles with different sizes. These particles are only interacting
via a hard-core repulsion, and hence temperature is not a relevant
parameter. We allow, however, pairs of particles to exchange volume
under the constraint that the total volume occupied by all particles
remains fixed. As a consequence the system has the freedom to explore
different size-distribution functions and relax to the ``optimal''
one. In the previous work \cite{Zhang:1999JCP} we found by computer
simulations that beyond a relative small pressure or density, some of
the particles became macroscopically large. From theoretical arguments
an upper bound to the pressure was obtained for fixed volume fraction,
above which a single size distribution is unstable. However,
simulations indicated a discrepancy with this result.  

The formation of macroscopic particles (in simulations of the size of
the simulation volume) beyond a critical value \cite{Sear:2001} is
in fact comparable to Bose-Einstein condensation, or alternatively as
a sort of sedimentation. In order to describe and simulate this
phenomenon in our system in a correct manner, one has to eliminate the
finite size effects that originate from the relative small number
(typically of the order 1000) of particles that is being used in
computer simulations. This could easily be solved by increasing the
number of particles by one or several orders of magnitude, but this is
seldom a preferred option. Moreover, in this particular case there
exists a much more elegant solution to the problem. 

The introduction of huge particles in a sea of smaller ones has two
effects. The first is the reduction of the available volume that can
be occupied by the smaller ones, and the second is the introduction of
a hard wall formed by the big particles. It turns out that it is the
second effect that caused the discrepancy between the simulation
results and the upper bound of the equation of state
\cite{Zhang:1999JCP}. The natural solution is therefore to eliminate
this particular effect, which can be done by extracting the big
particle(s) out of the simulation box.

We will assume that there is only a single big particle (aggregate)
with a volume $V_0 \geq 0$, but it can easily be extended to include
several aggregates, albeit their number with respect to the simulation
is not relevant. The total volume that can be occupied by the $N$
particles plus the aggregate, is denoted by $V_f = V_0 + N
\frac{\pi}{6} \langle \sigma^3 \rangle = N \frac{\pi}{6} \sigma_f^3$,
where $\langle \sigma^3 \rangle$ is the average diameter cubed of the
particles and $\sigma_f$ defines a natural length scale. The choice of
$\sigma_f$ in this formulation is based on the infinite dilute system
for which the size of the aggregate is zero. The volume that is
accessible to the particles is denoted by $V$. But since we have
extracted the aggregate out of the system the total thermodynamical
volume of the system is given by $V_T=V + V_0$.

On this system we have performed Monte Carlo (MC) simulations using
the isothermal-isobaric constant ($NPT$) ensemble. For a detailed
description on Monte Carlo simulations we refer the reader to
Ref. \cite{Book:Frenkel-Smit}; here we will only list the main
features. As positional order will play no role, we will assume the
simulation volume $V$ to be cubic and we will be using four different
types of moves.

The first and simplest move is randomly selecting a particle and
displacing it isotropically. The second type of move was introduced
before \cite{Zhang:1999JCP}. It selects randomly a pair of particles
and attempts to exchange a finite amount of volume, by adding this to
one, and subtracting it from the other particle. Hereafter their new
volumes are used in order to obtain the appropriate
diameters. Obviously, if this process led to a negative volume, the
move could not be accepted. Apart from that, both moves are accepted
provided that no overlap is caused by these changes. Moreover, the
amount of displacements and volume exchanges are drawn homogeneously
from an interval, of which the size is continously adjusted in order
to obtain an average acceptance per move between 35-50\%.

In the third move we allow the simulation box to shrink or expand
isotropically, in order for the system to equilibrate with respect to
the applied pressure. This is in principle also a standard move in MC
simulations using the $NPT$ ensemble. However, in this case it is
required to modify the acceptance criterion of this move. This is
caused by the fact that the thermodynamical volume $V_T$ and the
volume of the simulation box $V$, although related, are not identical.

The weight of any non-overlapping configuration in the $NPT$-ensemble
is proportional to $\exp(-\beta P V_T)$, where $\beta=1/(k_B T)$ the
inverse temperature. In addition to that, there is in MC-simulations a
correction factor $V^N$ due to the use of scaled coordinates in order
to facilitate the volume moves and periodic boundary conditions. As a
result the volume move, if no overlaps are produced, is accepted with
the probability
\begin{equation}
\mbox{Min}(1,\exp[\beta P (V_T^{(n)}-V_T^{(o)})-N
\log(V^{(n)}/V^{(o)})]),
\end{equation}
where the labels $n$ and $o$ refer to the new and old configuration
respectively.

The fourth type of move is the exchange of volume between a particle
and the aggregate. Although this seems to be equivalent to the volume
exchange between particles, there is a subtle difference. Provided
that no overlap is produced and both volumes remain positive this move
is allowed. However, if the aggregate changes its volume, this results
in a change of the available volume $V$ for the $N$ particles, while
the thermodynamic volume $V_T$ remains fixed. As a result this move
needs to be accepted with probability
\begin{equation}
\mbox{Min}(1,\exp[ -N \log(V^{(n)}/V^{(o)})]).
\end{equation}

We have performed a series of simulations on this system with $N=1000$
particles. In each simulation run we used $10^5$ sweeps, where in each
sweep on average we try $N$ particle moves, $N/2$ volume exchanges, 1
volume move, and 5 exchanges of volume between the aggregate and one
of the normal particles. Average quantities where obtained from a
single run after equilibrium was reached. The time required to reach
equilibrium strongly depends on the initial configuration. In
simulations where an equilibrium configuration of a slightly higher or
lower applied pressure is used, usually one run is sufficient. If
started from a monodisperse simple cubic lattice, one or several runs
are required, depending on the initial choice of the size of the
aggregate and volume of the simulation box.

\begin{center}
\begin{figure}[h]
\epsfig{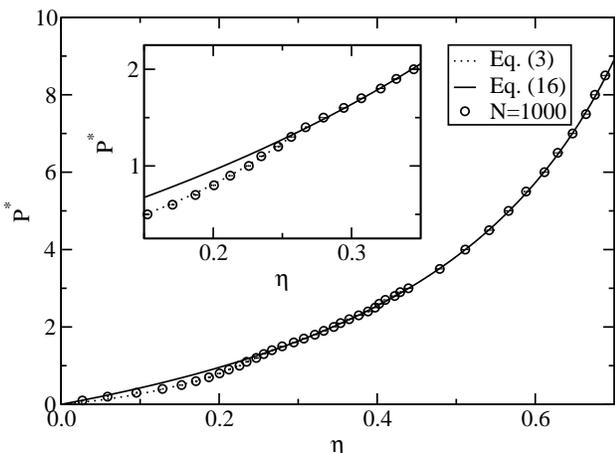}
\caption[a]{\label{fig:eos} 
Equation of state. The solid and dotted line are the theoretical
predictions above and below the transition respectively. The points are
obtained from MC simulations on a 1000 particle system. The inset
shows en enlargement of the transition point.}
\end{figure}
\end{center}

The results of the equation of state are shown in Fig.~\ref{fig:eos},
where the the reduced pressure $P^*=\beta P \sigma_f^3$ is plotted as
a function of the volume fraction $\eta=V_f/V_T$. The results obtained
from expansion and compression runs, as well as started from
monodisperse systems, all lead, within the estimated errors, to the
same results. Near the volume fraction 0.26, where the critical value
was found \cite{Zhang:1999JCP}, a change in slope can be observed.

\begin{center}
\begin{figure}[h]
\epsfig{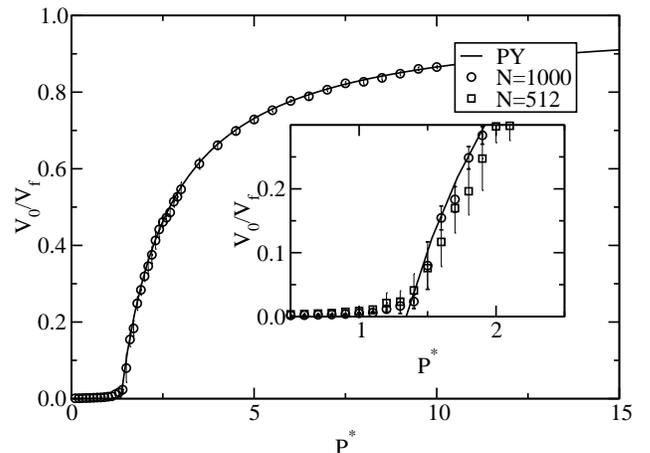}
\caption[a]{\label{fig:order}
The solid line is the theoretical prediction of the order parameter
$V_0/V_f$. The circles are results of MC-simulations on a 1000
particles system. The inset shows an enlargement of the area near the
phase transition and in addition simulation results of a 512 particle
system are indicated by squares.}
\end{figure}
\end{center}

A natural order parameter for this system is the relative amount of
fixed total volume of the particles found in the aggregate $V_0/V_f$
and is shown in Fig.~\ref{fig:order}. Comparison with simulations on
systems with less particles and the absence of hysteresis in the
equation of state suggests the possibility of a continuous phase
transition.

The simulation results below the critical value are consistent with
those obtained previously \cite{Zhang:1999JCP}. Above this value the
results are different. The presence of the aggregate seems to prevent
the formation of other macroscopic particles. Note that the difference
is not so much the fact that there is a aggregate, but the way it is
treated as an external buffer, thus eliminating the surface effects.

\section{Percus-Yevick Theory}
\label{sec:PY}
In order to describe the system under consideration theoretically, one
should realize that the presence of a thermodynamic volume $V_T$ as
well as an available volume $V$ for the particles should carefully be
taken into account. The system can be described by a polydisperse
system of $N$ hard spheres in a volume $V$, which is in contact with
an external bath, because in addition to the exchange of volume
between the particles themselves, this can also be done with this bath
or aggregate.

We will describe the polydisperse system within the Percus-Yevick
approximation of a polydisperse hard-sphere mixture
\cite{Salacuse:1982JCP}, from which the equation of state yields
\begin{equation}
\label{eq:PY}
\frac{\pi}{6} \beta P = \frac{\xi_0}{1-\xi_3} + \frac{3 \xi_1
\xi_2}{(1-\xi_3)^2} + \frac{3 \xi_2^3}{(1-\xi_3)^3},
\end{equation}
where $\xi_k$ is the $k$th moment of the particle-size distribution
function
\begin{equation}
\xi_k = \frac{\pi}{6} \rho \int dv W(v) \sigma^k = \frac{\eta}{1 -
\frac{V_0}{V_f}\eta} \frac{ \langle \sigma^k \rangle}{\sigma_f^3}.
\end{equation}
Here $\rho=N/V$ is the number density defined with respect to the
available volume, $\eta = V_f/V_T$ the volume fraction, and $W(v)$ is
the continuous particle-size distribution function, which due to the
type of interaction, is a function of the particle volume $v$. The
Helmholtz free energy functional $F(N,V_T,T)$ of this system, however,
is a function of the thermodynamic volume $V_T$
\begin{equation}
\frac{\beta F}{N} = \log(\rho \Lambda^3) - 1 + \int d v W(v) \log W(v)
+ \frac{\beta F^{ex}}{N},
\end{equation}
where the excess free energy per particle is given by
\begin{equation}
\frac{\beta F^{ex}}{N} = - \log(1-\xi_3) + \frac{3 \xi_1 \xi_2}{\xi_0
(1-\xi_3)} + \frac{3 \xi_2^3}{2 \xi_0 (1 - \xi_3)^2}.
\end{equation}
In order to obtain the equilibrium distribution, $W(v)$, we need to
minimize this free energy functional, taking into account that there
are two constraints which need to be satisfied. This is solved by
adding the following two terms to the free energy functional
\begin{equation}
- {\cal L}_0 \int d v W(v) - {\cal L}_3 \left( \int d v W(v) v +
  \frac{V_0}{N} \right),
\end{equation}
where ${\cal L}_0$ and ${\cal L}_3$ are Lagrange multipliers.
The first term is due to the requirement that the particle-size
distribution function should be normalized to unity, and the second is
that the combined volume of all particles and the aggregate is fixed
to be $V_f$.

The equilibrium particle-size distribution can now be obtained by
minimizing the free energy functional with respect to $W(v)$ and the
free parameter $V_0$. Stationarity with respect to changes in $W(v)$
leads to
\begin{eqnarray}
 && \log(W(v)) -\log(1-\xi_3) +\frac{3 \xi_2}{(1-\xi_3)} \sigma +
\\ && \left( \frac{3 \xi_1}{(1-\xi_3)} + \frac{9 \xi_2^2}{2
 (1-\xi_3)^2} \right) \sigma^2 + \frac{\pi}{6} \beta P \sigma^3 -
 {\cal L}_0 - {\cal L}_3 \frac{\pi}{6} \sigma^3 = 0, \nonumber
\end{eqnarray}
from which we can immediately derive the functional form of the
function $W(v)$ that will minimize the free energy
\begin{equation}
\label{eq:W}
W(v)=\exp \left( \sum_{i=0}^3 \alpha_i \sigma^i \right),
\end{equation}
where the coefficients $\alpha_i$ are determined by
\begin{equation}
\label{eq:alpha1}
\alpha_1 = - \frac{3 \xi_2}{(1-\xi_3)},
\end{equation}
\begin{equation}
\label{eq:alpha2}
\alpha_2 = - \left(\frac{3 \xi_1}{(1-\xi_3)} + \frac{9 \xi_2^2}{2
(1-\xi_3)^2} \right).
\end{equation}
The values of $\alpha_3$ and $\alpha_0$ are fixed by the constraint on
the total combined volume of the particles and the aggregate, and the
normalization of the particle size-distribution,
respectively. Formally, however, the coefficient $\alpha_3$ is given
by
\begin{equation}
\label{eq:alpha3}
\alpha_3 = - \frac{\pi}{6} (\beta P - {\cal L}_3).
\end{equation}

In addition to these equations we have to minimize with respect to the
volume of the aggregate. Note that the $\xi_i$ depend on the value of
$V_0$ through the number density. Therefore we obtain
\begin{equation}
\label{eq:stat}
\frac{\partial \beta F}{\partial V_0} = -\frac{\partial \beta
F}{\partial V_T} - {\cal L}_3 = \beta P - {\cal L}_3 = 0.
\end{equation}
This result does not automatically imply that $\alpha_3=0$. The reason
is that we have to minimize the free energy with respect to the volume
of the aggregate under the additional condition that the aggregate has
a non-negative volume. As a consequence this can result in the
possibility that the minimum is found for $V_0=0$ and the stationarity
equation (\ref{eq:stat}) not being satisfied. This leads to a natural
splitting of the behavior of this system. For low pressures the volume
of the aggregate will be zero, and the particle size-distribution is
the exponential of a third order polynomial in the particle
diameter. On increasing the pressure the coefficient $\alpha_3$, which
can only be non-positive in order to allow a proper normalization,
goes to zero. In this region the behavior is as explained before
\cite{Zhang:1999JCP}. According to the theoretical description in the
Percus-Yevick approximation, the coefficient $\alpha_3$ becomes zero
at volume fraction $\eta=0.260198$ pressure $P^*= 1.343442$. At this
point the value $V_0=0$ not only minimizes the free energy, but also
satisfies the stationarity equation (\ref{eq:stat}). For pressures
larger than this value the stationarity equation can always be
satisfied with a positive volume for the aggregate and hence the
coefficient $\alpha_3$ remains zero.

It turns out that this point is the location of a continuous
transition. Beyond this point the system relaxes by transferring
volume to the aggregate, and effectively scaling the polydisperse
system to a smaller size. If we denote by a script $c$ the values of
the critical system, this can be shown in the following way. If we
replace the coefficients $\alpha_k$ in the particle size-distribution
by $q^k \alpha_{ck}$, where $q$ is a positive number, and fix
$\alpha_0$ by the normalization to unity, it follows directly from the
special form (\ref{eq:W}) of $W(v)$ that the moments of the
size-distribution are given by $\langle \sigma^k \rangle = \langle
\sigma^k \rangle_c /q^k$. This results in
\begin{equation}
\alpha_1 = -q^3 \frac{3 \eta_c}{(1-\eta_c)} \frac{\langle \sigma^2
\rangle}{\sigma_f^3},
\end{equation}
and provided that $q$ is determined by
\begin{equation}
q^3 = \frac{\eta}{1-\eta} \frac{1-\eta_c}{\eta_c}
\end{equation}
this is a solution of Eq.~(\ref{eq:alpha1}). Applying this procedure
to $\alpha_2$ leads to the same expression for $q$. Therefore the
scaled size-distribution function leads to a minimum free energy,
albeit for a different applied pressure.

The pressure equation of state beyond the critical point also becomes
fairly simple now. If we rewrite the pressure (\ref{eq:PY}) using the
solutions (\ref{eq:alpha1}) and (\ref{eq:alpha2}), we obtain
\begin{equation}
\label{eq:PY-high}
\frac{\pi}{6} P^* = \frac{\xi_0 - \frac{1}{3}\alpha_1 \xi_1 -
\frac{2}{3} \alpha_2 \xi_2}{1-\xi_3} \sigma_f^3 = \frac{2
\eta}{1-\eta},
\end{equation}
where we used that if $\alpha_3=0$ the functional form of $W(v)$ leads
to the identity $3 \langle \sigma^0 \rangle + \alpha_1 \langle \sigma
\rangle + 2 \alpha_2 \langle \sigma^2 \rangle = 0$. We can also derive
a simple expression for the order parameter $V_0/V_f$ of the system
\begin{equation} 
\label{eq:order}
\frac{V_0}{V_f} = 1 - \frac{\langle \sigma^3 \rangle}{\sigma_f^3} = 1
-\frac{P^*_c}{P^*}.
\end{equation}
The curves of the equation of state and the order parameter are shown
in Figs.~\ref{fig:eos} and \ref{fig:order} respectively, and show a
perfect agreement with the simulation results. In Fig.~\ref{fig:eos}
the dotted line represents the equation of state below the transition,
i.e.\ Eq.~(\ref{eq:PY}), while the solid line is the one above as given
by Eq.~(\ref{eq:PY-high}). The later one is extended to the zero
density limit in order to visualize the difference between both
branches.

\begin{center}
\begin{figure}[h]
\epsfig{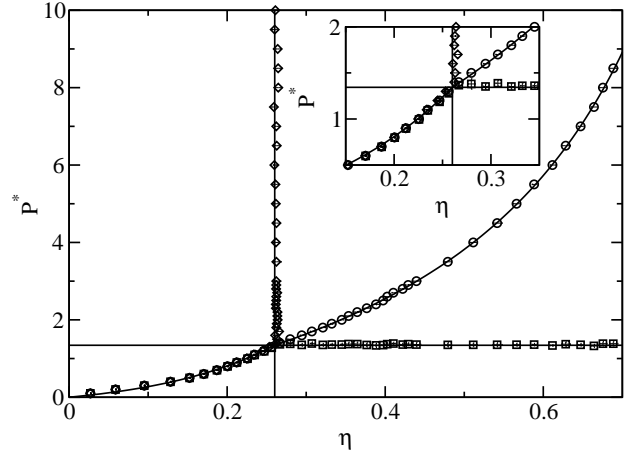}
\caption[a]{\label{fig:transition}
Comparison of the simulation results for a 1000 particles system with
the theoretical predictions. The circles represent the applied
pressure $P^*$ as function of the volume fraction $\eta$. The squares
the local pressure $P^* \langle \sigma^3 \rangle /\sigma_f^3$ as
function of $\eta$, and the diamonds the applied pressure $P^*$ as
function of the local volume fraction $N \langle v \rangle/V$.}
\end{figure}
\end{center}

From these results two additional illustrative properties can be
derived for the system beyond the point of the transition. Using that
the particle size-distribution only changes according to a dimensional
scaling or from the order parameter (\ref{eq:order}), we find that
\begin{equation}
P \langle \sigma^3 \rangle = P_c \sigma_f^3.
\end{equation}
This simply states that the system becomes scale invariant, with a
constant pressure if it is defined in terms of the average particle
volume. The other result is that $\xi_3$, which can be identified with
the local volume fraction of the particles, remains fixed, i.e. $\xi_3
= \eta_c$. These results can be used in order to determine the point
of transition from simulations. This is illustrated in
Fig.~\ref{fig:transition}, where $P^* \langle \sigma^3
\rangle/\sigma_f^3$ as a function of $\eta$, and the applied pressure
$P^*$ as a function of $\xi_3$ are shown, as measured in our
MC-simulations. According to these simulations the transition is found
at a volume fraction $\eta=0.2623 \pm 0.0003$ and pressure $P^*=1.360
\pm 0.003$.

\begin{center}
\begin{figure}[h]
\epsfig{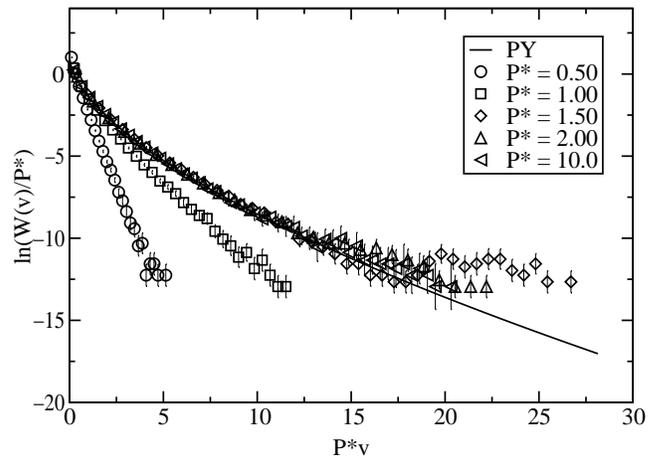}
\caption[a]{\label{fig:sizedist}
The particles size-distributions are shown by plotting
$\mbox{ln}(W(v)/P^*)$ as a function of $P^* v$. Due to this scaling
all distributions above the transition should coincide, without using
the actual values of the pressure and volume fraction at the
transition. The solid curve is the predicted curve by the
Percus-Yevick approach.}
\end{figure}
\end{center}

The prediction of the particle size-distribution $W(v)$ is compared
with the simulation results in Fig.\ref{fig:sizedist}. By plotting
$\mbox{ln}(W(v)/P^*)$ as a function of $P^* v$, the distributions
above the transition coincide and agree perfectly with the predicted
curve from Percus-Yevick. The discrepancy at higher values for $P^* v$
is merely a consequence of an insufficient sampling of the largest
particles in the system. For $P^* = 1.50$, there is the additional
effect that this is close to the transition, where the system tends to
probe larger particles.

It turns out that within the Percus-Yevick approximation the system
shows a continuous phase transition, which is characterized by the
non-zero volume of a aggregate. In fact it shows some similarities to
Bose-Einstein condensation, because what actually happens is that the
interaction of microscopic particles leads to the formation of a
macroscopic aggregate \cite{Sear:2001}.

\section{BMCSL-theory}
\label{sec:BMCSL}
It is a well known fact that the Percus-Yevick equation of state in a
monodisperse hard sphere system can be improved by combining it with
the virial equation of state. This results in the Carnahan-Starling
equation of state \cite{Carnahan:1969JCP}. A similar approach can be
made for polydisperse sphere mixtures, and the resulting heuristic
equation of state is due to Boubl\'{\i}k, Mansoori, Carnahan,
Starling, and Leland (BMCSL) \cite{Boublik:1970JCP,Mansoori:1971JCP}
\begin{equation}
\label{eq:MCSL}
\frac{\pi}{6} \beta P = \frac{\xi_0}{1-\xi_3} + \frac{3 \xi_1
\xi_2^2}{(1-\xi_3)^2} + \frac{3 \xi_2^3}{(1-\xi_3)^3} - \frac{\xi_3
\xi_2^3}{(1-\xi_3)^3}.
\end{equation}
A straightforward integration of this equation of state shows that the
excess free energy up to a constant is given by
\cite{Salacuse:1982JCP}
\begin{eqnarray}
\frac{ \beta F^{ex}}{N}&=&\left(\frac{\xi_2^3}{\xi_0 \xi_3^2}-1\right)
\ln(1-\xi_3) \nonumber \\
&&+\frac{3 \xi_1 \xi_2}{\xi_0(1-
\xi_3)}+\frac{\xi_2^3}{\xi_0 \xi_3(1-\xi_3)^2}.
\end{eqnarray}
The additional constant is independent of the volume, but could in
principle depend on the moments of the size-distribution, without
changing the equation of state. Such a constant would be relevant for
this particular model, since the size-distribution of the system can
be changed, and therefore an additional change in free energy can be
produced. From general applicability of the free energy functional to
different models, however, it can be shown that this contribution
necessarily vanishes.

We now can apply the same procedure as was used in the case of the
Percus-Yevick approximation. The functional derivative of the free
energy with respect to $W(v)$ leads to the same functional form
(\ref{eq:W}) as before, but the stationarity equations on the
coefficients $\alpha_i$ need to be modified
\begin{equation}
\label{eq:alpha1-CS}
\alpha_1 = - \frac{3 \xi_2}{(1-\xi_3)},
\end{equation}
\begin{equation}
\label{eq:alpha2-CS}
\alpha_2 = - \frac{3 \xi_1}{(1-\xi_3)} - \frac{3 \xi_2^2}{\xi_3
(1-\xi_3)^2} - \frac{3 \xi_2^2}{\xi_3^2} \log(1-\xi_3),
\end{equation}
\begin{eqnarray}
\label{eq:alpha3-CS}
\alpha_3 &=& - \frac{\pi}{6} (\beta P - {\cal L}_3) \nonumber \\
&&+ \left(
\frac{\xi_2}{\xi_3} \right)^3 \left[ 2 \log(1-\xi_3) +
\frac{\xi_3(2-\xi_3)}{(1-\xi_3)} \right].
\end{eqnarray}
The derivative of the free energy with respect to the volume of the
aggregate leads again to
\begin{equation}
\label{eq:stat-CS}
\frac{\partial \beta F}{\partial V_0} = -\frac{\partial \beta
F}{\partial V_T} - {\cal L}_3 = (\beta P - {\cal L}_3) = 0.
\end{equation}
The second term on the right hand side of Eq.~(\ref{eq:alpha3-CS}) is,
except in the zero density limit, always positive. As a consequence
Eq.~(\ref{eq:stat-CS}) can never be satisfied, because such a solution
would lead to a positive $\alpha_3$ and hence a non-normalizable
size-distribution function. The minimum free energy for given volume
fraction will therefore not be a solution of all the stationarity
equations simultaneously. 

\begin{center}
\begin{figure}[h]
\epsfig{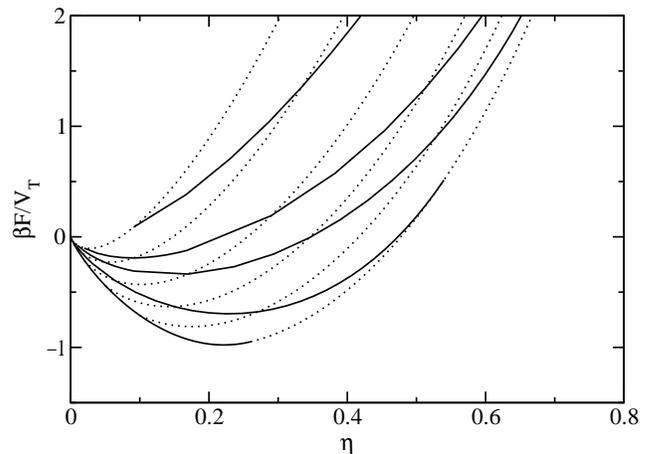}
\caption[a]{\label{fig:free} 
The Helmholtz free energy per volume as a function of the volume
fraction for fixed order parameter $V_0/V_f$, i.e. constant size of the
aggregate (solid curves) and for scaled distribution functions (dotted
curves). Both sets are similar but use only a different parameter
for characterization.}
\end{figure}
\end{center}

Since the stationarity equations (\ref{eq:alpha1-CS}) -
(\ref{eq:alpha3-CS}) are directly related to the equations for chemical
equilibrium between the different particle sizes, they need to be
fulfilled. If we fixed the volume of the aggregate $V_0$, these
equations would be sufficient to obtain the unique distribution
function that minimizes the free energy. In Fig.~\ref{fig:free} we
have plotted the free energy per volume for several fixed values of
$V_0$ (solid curves). On these curves the stationarity equations are
satisfied, hence the value of $V_0$ that leads to the minimum free
energy corresponds to the equilibrium situation. These curves are
bounded by an lower and upper limit of the volume fraction. The lower
limit corresponds to the case where $\alpha_1 = \alpha_2 = 0$ and the
upper limit corresponds to the case where $\alpha_3=0$. Outside this
interval no self-consistent solution of the stationarity equations can
be obtained.  

Similar to what we found in the Percus-Yevick approach, we can scale
the particle size-distribution function such that we obtain a solution
to the stationarity equations for a different volume fraction by
replacing $\alpha_k$ by $q^k \alpha_k$ and fixing the normalization
constant. The free energy per volume for several of these scaled
solutions is also shown in Fig.~\ref{fig:free} as dotted curves. Both
sets of curves are similar, but use a different parameter to
describe them.

The equilibrium situation of our model is the one that corresponds to
the lowest free energy for given volume fraction and will correspond
to the envelope of the set of curves in Fig.~\ref{fig:free}. One can
show that we have two regions, which is illustrated by this figure.
Below the transition the equilibrium situation corresponds to the
curve for which $V_0=0$, i.e. without a macroscopic 
aggregate. Above the transition we will follow the branch of a scaled
solution, emerging from the end point of this curve determined by
$\alpha_3=0$. The transition point is therefore obtained by
self-consistently solving the case for $V_0=\alpha_3=0$ and leads to a
volume fraction $\eta=0.262611$ and pressure $P^*=1.356275$.

Actually the same reasoning could have been used for the Percus-Yevick
approach. But since for that case all stationarity equations could be
solved simultaneously, there was no need to follow this line of
argumentation. 

The numerical predictions of the transition point for the BMCSL
approach are slightly better in agreement with the simulation results
than the results from Percus-Yevick. The difference between both
predictions is however too small in order for it to be visible in any
of the figures. In principle also the size-distribution function can
be used to distinguish both theories, however, in order to do so the
number of particles should be increased several orders of
magnitudes in order to obtain sufficient statistics for the larger
particles and to make the difference in the tail of the distribution
visible.

\section{Discussion}
\label{sec:discussion}
We have demonstrated how to obtain a self-consistent theory for a
polydisperse hard sphere system, where particles are able to change
their size by exchanging volume. Within the Percus-Yevick
approximation this system shows a continuous phase transition, in
which the microscopic interactions lead to the formation of a
macroscopic aggregate, a behavior similar to Bose-Einstein
condensation. The theoretical results have been compared with a new
series of computer simulations and show an excellent agreement,
because in the new treatment we have eliminated the finite size effect
that previously led to discrepancies between theory and simulation.

This peculiar system, which at best can be considered to be an
extremely idealized version of spherical micelles, exposes an unusual
behavior in the BMCSL-approach. The equilibrium state of the system is
one which does not satisfies the stationarity equations but is
determined by the boundary conditions on the problem, i.e. a
normalizable particle size-distribution. Apart from that the
difference between both approximations is minor and in order to
determine which one is better, simulations would be required with
several orders of magnitude more particles.

\section*{Acknowledgments}
R.B. acknowledges the financial support of the EU through the
Marie Curie Individual Fellowship Program (contract no.~%
HPMF-CT-1999-00100). J.A.C.\ acknowledges financial support
of Ministerio de Ciencia y Tecnolog\'{\i}a (Spain) through the
project no.~BFM2000-0004.

\end{multicols}


\begin{thebibliography}{1}

\bibitem{Sear:98EPL}
R.~P. Sear, Europhys. Lett. {\bf 44}, 531 (1998).

\bibitem{Bartlett:98JCP}
P. Bartlett, J. Chem. Phys. {\bf 109}, 10970 (1998).

\bibitem{Cuesta:99EPL}
J.~A. Cuesta, Europhys. Lett. {\bf 46}, 197 (1999).

\bibitem{Warren:99EPL}
P.~B. Warren, Europhys. Lett. {\bf 46}, 295 (1999).

\bibitem{Bartlett:99PRL}
P. Bartlett and P.~B. Warren, Phys. Rev. Lett. {\bf 82}, 1979 (1999).

\bibitem{Bates:1998JCP}
M.~A. Bates and D. Frenkel, J. Chem. Phys. {\bf 109}, 6193 (1998).

\bibitem{Gelbart:1994book}
W.~M. Gelbart, A. Ben-Shaul, and D. Roux, eds., {\em Micelles, 
Microemulsions and Monolayers} (Springer-Verlag, New York, 1994).

\bibitem{Boden:1986CPL}
N. Boden, R.~J. Bushby, C. Hardy, and F. Sixl, Chem. Phys. Lett.
{\bf 123}, 359 (1986).

\bibitem{Boden:1987JPC}
N. Boden, S.~A. Corne, and K.~W. Jolley, J. Phys. Chem. {\bf 91},
4092 (1987).

\bibitem{Leaver:1994JPII}
M.~S. Leaver, U. Olsson, H. Wennerstrom, and R. Strey, J. Phys. II
{\bf 4}, 515 (1994).

\bibitem{Vollmer:1997JCP}
D. Vollmer, R. Strey, and J. Vollmer, J. Chem. Phys. {\bf 107},
3619 (1997).

\bibitem{Petschek:1986PRA}
R.~J. Petscheck, P. Pfeuty, and J.~C. Wheeler, Phys. Rev. A {\bf 34},
2391 (1986).

\bibitem{Sollich:1998PRL}
P. Sollich and M.~E. Cates, Phys. Rev. Lett. {\bf 80}, 1365 (1998).

\bibitem{Warren:1998PRL}
P.~B. Warren, Phys. Rev. Lett. {\bf 80}, 1369 (1998).

\bibitem{Sollich:2001ACP} 
P. Sollich, P.~B. Warren, and M.~E. Cates, Adv. Chem. Phys. 
{\bf 116}, 265 (2001).

\bibitem{Clarke:2000JCP}
N. Clarke, J.~A. Cuesta, R. Sear, P. Sollich, and A. Speranza,
J. Chem. Phys. {\bf 113}, 5817 (2000).

\bibitem{Zhang:1999JCP}
J. Zhang, R. Blaak, E. Trizac, J.~A. Cuesta, and D. Frenkel, 
J. Chem. Phys. {\bf 110},  5318  (1999). 

\bibitem{Blaak:2000JCP}
R. Blaak, J. Chem. Phys. {\bf 112},  9041  (2000).

\bibitem{Sear:2001}
R.~P. Sear and J.~A. Cuesta, cond-mat/0012256 (2000).

\bibitem{Frenkel}
D.~Frenkel, Private Communication.

\bibitem{Book:Frenkel-Smit}
D. Frenkel and B. Smit, {\em Understanding Molecular Simulation. From
  Algorithms to Applications} (Academic Press, Boston, 1996).

\bibitem{Salacuse:1982JCP}
J.~J. Salacuse and G. Stell, J. Chem. Phys. {\bf 77},  3714  (1982).

\bibitem{Carnahan:1969JCP}
N.~F. Carnahan and K.~E. Starling, J. Chem. Phys. {\bf 51},  635  (1969).

\bibitem{Boublik:1970JCP}
T. Boubl\'{\i}k, J. Chem. Phys. {\bf 53},  471  (1970).

\bibitem{Mansoori:1971JCP}
G.~A. Mansoori, N.~F. Carnahan, K.~E. Starling, and T.~W. {Leland, Jr.}, J.
  Chem. Phys. {\bf 54},  1523  (1971).

\end{thebibliography}
\end{document}